\documentclass{kapproc} 
\renewcommand{\vec}[1]{\mathbf{#1}}
\def\be{\begin{equation}}
\def\ee{\end{equation}}
\def\bea{\begin{eqnarray}}
\def\eea{\end{eqnarray}}
\usepackage[dvips]{graphicx}
\usepackage{procps}

\let\footnotetext\savefootnotetext
\let\footnoterule\savefootnoterule
\upperandlowercase

\nochapequationnumber 

\normallatexbib

\begin{document}


\articletitle[Phase Transitions in High Energy\\
Heavy-Ion Collisions]
{Phase Transitions in High Energy\\ Heavy-Ion Collisions}

\chaptitlerunninghead{Phase Transitions in ... }

\author{%
L.P. Csernai$^{1,2}$, Andrea Anderlik$^1$, Cs. Anderlik$^1$, A. Keranen$^{3,4}$,\\
V.K. Magas$^5$, J. Manninen$^4$, E. Moln\'ar$^1$, \'Agnes Ny\'{\i}ri$^1$, B.R. Schlei$^6$,\\
D.D. Strottman$^6$ and K. Tamosiunas$^1$
}
\smallskip

\affil{%
$^{1}$
Section for Theoretical and Computational Physics, and Bergen Centre for
Computational\\
\hskip 0.3cm Science - Unifob, University of Bergen, All\'egaten 55, 5007 Bergen, Norway\\
$^2$ Res. Inst. for Particle and Nuclear Physics, POB 49, 1525 Budapest, Hungary\\
$^3$ INFN, Sezione di Firenze, Via Sansone 1, 50019 Sesto F. - Florence, Italy\\
$^4$ Department of Physics, POBox 3000, 90014 Univeristy of Oulu, Finland\\
$^5$ Departamento de Fisica Teorica, Dr. Moliner 50, 46100 Burjassot, Valencia, Spain\\
$^6$ Los Alamos National Laboratory, Theory Division, Los Alamos, 87545 NM, USA
}

\begin{abstract}
Modelling Quark-Gluon Plasma formation and decay in high energy
heavy ion reactions is presented in a framework of a multi-module
setup. The collective features, governing the equlibrated fluid dynamical
stages of the model are emphasized. Flow effects formed from
the initial conditions are discussed. Particular attention is given to the
improvement of the final hadronization and freeze-out part of the reaction
which has strong effects on the observables.
\end{abstract}

\section{Introduction}
Phase Transitions imply that we have different phases of the matter
we deal with, and it has an Equation of State (EoS). We can have an EoS
if and only if the matter is in statistical equilibrium, or at least
close to it. When this condition is satisfied we can apply a fluid-
and thermodynamical description for the dynamics of the system.
In high energy heavy ion reactions we observe many thousand particles
produced in the reaction, so, we have all reasons to assume that in a
good part of the reaction the conditions of the local equilibrium and
continuum like behaviour are satisfied.

The initial and final stages are, on the other hand, obviously not in
statistically equilibrated states, and must be described separately,
in other theoretical approaches. The different approaches, as they
describe different space-time (ST) domains and the corresponding
approaches can be matched to each other across ST hyper-surfaces or
across some transitional layers or fronts. The choice of realistic
models at each stage of the collision, as well as the correct coupling of
the different stages or calculational modules are vital for a
reliable reaction model. The most basic requirements are that conserved
quantities should stay conserved when modules are interfaced to each other,
and entropy should not decrease.

For the theoretical description and understanding of high energy
heavy ion collisions it is important to mention that while our system is
sufficiently large to assume statistical equilibrium, it is not a
macroscopic system with Avogadro number of particles. We deal with a
small, mesoscopic system, so discontinuities of the thermodynamical
quantities, will not be observable, but at the same time the changes
are sudden and rapid enough that they can be recognized.

The other basic feature of mesoscopic systems is also important to us:
in such systems fluctuations are not negligible but dominant, these can be
observed and these may help to identify the properties of the phase
transition.

Dividing the reaction model into standardized modules and interfaces leads
to additional advantages. Modules can be replaced easily, research groups
can collaborate better and the GRIDS of High Performance Computers can
increase the speed of complete reaction simulation by about an order of
magnitude.

\section{Initial Stage Module}

Many fluid dynamical approaches use simple parametrizations for the
initial state of the hydrodynamic stage with a few parameters. These
can then be fitted or adjusted by comparing experimental data and
model predictions.

A more physical approach is to start from a dynamical, pre-equilibrium
model, e.g. Parton Cascade Model (PCM), or Collective (or Coherent)
Yang-Mills model, or different versions of flux-tube models.

Our recent works are based on a latter type of approach \cite{GC86},
which was implemented in a Fire-Streak geometry streak by streak and
upgraded to satisfy energy, momentum and baryon charge conservations
exactly at given finite energies \cite{MCS01,MCSe02}. The effective
string-tension was different for each streak, stretching in the
beam direction, so that central streaks with more color (and baryon) charges
at their two ends had bigger string-tension and expanded less, than peripheral
streaks. The expansion of the streaks was {\it assumed} to last until the
expansion has stopped. Yo-yo motion, as known from the {\it Lund-model} was
not assumed.

In the 1st version of the model at full stopping of string
expansion, the baryon charge was distributed uniformly along each
streak. This created an initial state for the hydrodynamic stage
where each streak was stopped in its own CM frame, and had uniform
baryon and energy density distribution. The sharp forward and backward
end led, however, to a development of a large, Fwd/Bwd density peaks.
\cite{MCSb01}

Thus, we concluded that the assumption of uniform distribution and sharp
cuts at the ends is overly simplified and has to be corrected. So, a
Fwd/Bwd expansion was added to the model  \cite{MCSe02},
which smoothed the Fwd/Bwd
ends, and eliminated the artificial large, Fwd/Bwd density peaks.

Our calculations show that such a tilted initial state leads to the
creation of of the third flow component \cite{CR99}, peaking at rapidities
$|y| \approx 0.75$ \cite{MCSb01}. Recent, STAR $v_1$ data \cite{Star-v1}
indicate that our {\it assumption} that the
string expansion lasts until full stopping of each streak, may
also be too simple and local equilibration may be achieved earlier,
i.e. before the full uniform stopping of a streak. We did not
explicitly calculate dissipative processes, some friction within and
among the expanding streaks is certainly present and experiments seem
to indicate that this friction is stronger.

\section{Relativistic Fluid Dynamics}

When parts of our system reach local statistical equilibrium in the
ST, following a pre-equilibrium development, we can describe our system
using Computational FLuid Dynamics (CFD) and an Equation of State.
The equations of Fluid Dynamics (FD) can be most simply
derived from the Boltzmann Transport Equation (BTE).

The relativistic BTE below, describes the time evolution of the single particle
distribution function based on the assumptions that just two particle
collisions are considered (binary collisions), the number of binary collisions
at $x$ is proportional to $f(x,p_{1})\times$ $f(x,p_{2})$ and $f(x,p)$ is a
smoothly varying function compared to mean free path. So, one can show
that BTE has the form [see Chapter 3 of ref. \cite{LasBook}] :%

\[
p_{k}^{\mu}f_{k,\mu}=\sum_{l=1}^{N}C_{kl}(x,p_{x})
\]

where $C$ is the collision integral, and $k$ and $l$ stand for different
particle species or particle components. Then, using microscopic conservation
laws in the collision integral, from the BTE we can derive the differential
form of conservation laws [see section 3.6 of ref. \cite{LasBook}]:%

\bea
N_{,\mu}^{\mu}  & =&\sum_{k=1}^{N}N_{k,\mu}^{\mu}=0\rm{
\ \ \ \ :conservation\ of\ particle\ number} \label{Cons-law1}\\
Q_{,\mu}^{\mu}  & =&\sum_{k=1}^{N}Q_{k,\mu}^{\mu}=0\rm{
\ \ \ :conservation\ of\ charge}\\
T_{,\mu}^{\mu\nu}  & =&\sum_{k=1}^{N}T_{k,\mu}^{\mu\nu}=0\rm{
\ \ :conservation\ of\ energy\ and\ momentum }%
\label{Cons-law3}
\eea

where $_{,\mu}$ denotes $\partial / \partial \mu $ and there is a summation
for indexes occurring twice. These equations are valid for any distribution
satisfying the BTE.  These equations can be
used if we know the solution and then
we can evaluate the conserved quantities.

\subsection{Perfect and Viscous Fluids}

Equations (\ref{Cons-law1}-\ref{Cons-law3}) are also the equations of FD.
But these equations must not be considered
as consequences of BTE, rather these equations are {\bf postulated},
as the differential forms of conservation laws. This can be done because
such conservation laws can be derived from other
theoretical approaches also.

These equations are not a closed set, because the energy
momentum tensor and the particle four current should also be defined.
In the Eulerian or perfect fluid dynamics we postulate the form of
energy-momentum tensor, and an EoS, $P=P(e,n)$, must be given.
This provides a closed set of partial
differential equations that is solvable. Any EoS,
which is consistent with thermodynamics, can be
used. One does not have to assume dilute systems with binary collisions,
or the assumption of molecular chaos. The only requirement is the existence
of local equilibrium, otherwise we would not have an EoS.

In the case of Navier-Stokes fluid dynamics, we assume that the energy
momentum tensor also contains the dissipative part $T^{\mu\nu (1)}$,
which allows for small, first order deviations
from local equilibrium.  Thus, to have a solution we need
not only the EoS, but also the
transport coefficients that occur in the dissipative part of the
energy-momentum tensor.

Boltzmann's H-theorem implies that
if irreversible processes are present the entropy increases. In
equilibrium the distribution is constant, the entropy is also constant, but it
reached its maximum value.

It is important to know that this can be seen from  fluid dynamics also.
In perfect fluid dynamics the flow is adiabatic, no entropy is produced.
This can be proven using the equations of perfect fluid dynamics and
standard thermodynamical relations.
Entropy production is strictly, and quantitatively connected to
dissipative processes.
Perfect flow is adiabatic even in the presence of phase transitions, if
the two phases are in thermal, mechanical ($P$), chemical and phase
equilibrium all the time! On the other hand if the phase transition
deviates from equilibrium this leads to entropy production. See
Assignment 9.4.a in ref. \cite{LasBook}.

As shock waves, detonations, deflagrations, are idealizations of very sharp
or very rapid processes as discontinuities, the above conclusions are
valid for discontinuities also. Nevertheless, frequently, dissipative
processes can be neglected for most of the flow except the sharp or rapid
changes. Thus, as an end effect the dissipative processes are frequently
localized in sharp fronts or in hypersurfaces of discontinuities.
One should not, however, forget that dissipation is due to transport
properties and characteristic constants of phase transition dynamics, even
if it happens in "discontinuities".

\subsection{Equations of Perfect Fluid Dynamics}

We assume that our system is not homogeneous, but the gradients are small, so
local distribution can be an equilibrium distribution, e.g. a J\"uttner
distribution.
Now the local $n(x), e(x), s(x)$ and $P=P(n,e)$ are
known, and we assume that in LR, $T^{\mu\nu}$ is diagonal. Now
using the conservation laws with
introducing the apparent density as:%

\[
N \equiv n\gamma\ ,
\]

the continuity equation takes the familiar form%

\[
(\partial_{t}+\vec{v}\ \rm{grad})\ N
= - M \ \rm{div}\ \vec{v}\ ,
\]

similarly introducing%

\bea
\vec{M}  & \equiv & T^{0i}=w\ \gamma^{2}\vec{v}\\
\varepsilon & \equiv & T^{00}=(e+P\vec{v}^{2})\gamma^{2}\ ,%
\eea

the energy and momentum conservation will take the form%

\bea
(\partial_{t}+\vec{v}\ \rm{grad})\vec{M}  &
= -\vec{M}(\rm{div} \vec{v})- \rm{grad}\ P\\
(\partial_{t}+\vec{v}\ \rm{grad})\varepsilon & =-\varepsilon\
\rm{div} \vec{v}-\rm{div} (P\vec{v})\ .%
\eea

Thus to solve the equations of relativistic fluid dynamics we have to solve
the same partial differential equations as in the non-relativistic case.
We cannot immediately apply the EoS to obtain the solution, as the
EoS is given in terms of invariant scalars. So, we have to solve in addition
the set of algebraic equations above, which connect $N$, $\vec{M}$,
and $\varepsilon$ to $n$, $e$, $P$ and $\vec{v}$, for every fluid
cell at every time step.

Important to note that the relativistic treatment is necessary not only
at high velocities (i.e. if $v \sim c$) but also at low velocities when
the pressure is not negligible compared to the energy density (i.e.,
$ P \sim e$) as $w = e + P$ appears in the relativistic Euler equation!
This is frequently the situation for ultra-relativistic gases, like the
Stefan-Boltzmann or photon gas (or radiation pressure) where $P=e/3$.
This is actually the indication of the fact that the constituents of the
matter move with velocity $c$ or close to it.

In principle the perfect fluid dynamics is absolutely unstable. Any
spontaneous deviation from an exact solution, will grow exponentially, as
their energy cannot be dissipated away \cite{LLHCh3}. In other words
the Reynolds number tends to infinity and the system tends to turbulence.
Viscosity is needed to stabilize the flow.

\begin{figure}[ht]
\centering
\includegraphics[width=10cm, height = 7.5cm]{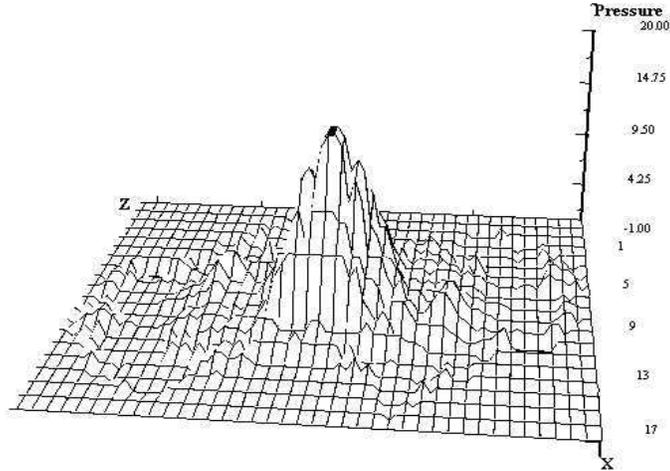}
\caption{
The pressure, in units of [GeV/fm$^3$], in the reaction plane at a
late stage (13 fm/c) of a 65+65 Gev/nucl. Au+Au collision, at b=0.5 b-max,
using the Bag Model EoS for QGP. The outer, supercooled regions have
negative pressure, while in the very center it is still large. The directed
flow, is visible as the outside, negative pressure regions in the
upper-right and lower-left corners, close to the beam, $Z-$, direction.
The third flow component \cite{CR99}
is shown by a smaller negative pressure peak
at the lower side of the plot, slightly right of the middle. The section
of the reaction plane, [x,z], shown is 12.1 $\times$ 9.6 fm.
}
\label{P_polt-A}
\end{figure}

There are several numerical methods to obtain a solution. In all cases
the solution is discretized in some way and a coarse graining is introduced.
This coarse graining automatically leads to dissipative and transport
properties, and the parameters of these transport processes can be determined
by numerical experiments.  The result is that all
CFD models are dissipative or viscous, which is an advantage as the
these are usually stable. By decreasing the cell size, the dissipative
effects can be decreased in CFD, to a limit when instabilities, or turbulence
occurs. As a side effect, CFD models lead to entropy production, as part of
the kinetic energy is dissipated away due to the coarse graining or smoothing.

\section{Kinetic Theory of Particle Formation, Escape and Freeze-out}

Most frequently relativistic kinetic theory describes
the time, $t$, development of the single particle distribution function,
                    $f(p;x)=f(p,\vec{r},t)$,
of conserved particles in the 6-dimensional
Phase-Space (PS), $[\vec{p},\vec{r}]$ and time. Then the total particle
number does not change with time. We assume (usually) that the particles are
on mass shell, so the 4-momentum has 3 independent variables. The
phase space distribution is a density in the 6 dimensional PS and
it is an invariant scalar, because a Lorentz boost contracts the
system in the configuration space but elongates it in the momentum
space.

If we want to describe a particle source (or drain) from a matter,
the total number of newly crated particles can be given as
       $$ N_{new} = \int d^4p \ d^4x\  g(p,x) , $$
where $g(p,x)$ is an invariant scalar density in the 4-dimensional
PS and 4-dimensional space-time (ST).  If the particles are on mass shell the
momentum space integral can be reduced to
\be
       N_{new} = \int \frac{d^3p}{p^0} \ d^4x\  h(p,x) ,
       \label{NewH}
\ee
where $h(p,x)$ is also an invariant scalar density, as
$\frac{d^3p}{p^0}$
is also one.

The particle formation or source density $h(p,x)$ from a matter, can
be characterized by 3 factors: \\
PS distribution,
$f(p,\vec{r},t)$,\\
PS emission probability,
$P_{esc-PS}(p;x)$,\\
ST emission probability,
$P_{esc-ST}(x;p)$.\\
Here $f$ can be an arbitrary distribution, but we usually discuss
matter in (or close to) statistical equilibrium, with known
PS distribution. $P_{esc-PS}$ is most frequently assumed to be unity, while
$P_{esc-ST}$ is the ST distribution if the source (e.g. a
4-dimensional Gaussian in the ST). The experimental analysis of two
particle correlations from heavy ion reactions, is based on these
very simple assumptions in most cases.

These are very strong assumptions. It is not supported by experiments that
the emission probability is isotrope and homogeneous in the PS, or neither
that the ST distribution of the source is Gaussian.

\subsection{Non-isotropic particle sources}

Not only in heavy ion reactions but in any dynamical process,
particle creation, (or condensation) happens mostly in a directed way: the
phenomenon propagates into some direction, i.e. it happens in
some (space-like) layer or front (like detonations, deflagrations, shocks,
condensation waves or freeze-out). The reason is that neighbouring
regions in the front may interact, minimize the energy of the front
by evening it out, provide energy to neigbouring regions to
exceed the threshold conditions. Even in relativistic processes, that
are time-like, (have time-like normal) and the neighbouring points
of a front cannot be in causal connection, in the LR frame of the matter
the dynamical process may and frequently has a direction. (See the example
in ref. \cite{Cs87}.)

These fronts or layers are not necessarily narrow, but they have
a characteristic direction (or normal, $d^3\sigma^\mu$), thus the ST
integral, $d^4x$, can be converted into
     $$ d^4 x  \longrightarrow  ds^\mu\ d^3\sigma_\mu .$$
The front can actually be directed time-like,
$d^3\sigma^\mu \ d^3\sigma_\mu = + 1$,
or space-like,
$d^3\sigma^\mu \ d^3\sigma_\mu = - 1$.

An example for emission distributed in the ST, but being time-directed can be
described by the expression:
\be
    N_{new} = \int d^3p \int d\tau  d^3\sigma_\mu\ \
    f_{eq}(p,x)\ \
    \frac{C_1 p^\mu}{p^0}\  \Theta(p^\nu\ d^3\sigma_\nu)\ \
    G\left( \tau-\tau_{FO}(x) \right)
    \label{eq11}
\ee
where
\be
  \frac{C_1 p^\mu}{p^0}\ \Theta(p^\nu\ d^3\sigma_\nu)\
\label{EmiProb}
\ee
is the PS emission probability, where the step function eliminates
the possibility of emission of particles, in the direction opposite
to the normal of the front (for time-like normals it is always  $+1$),
and the term,
$\frac{p^\mu}{p^0} = ( 1, \vec{v} )$,
together with the 6- dimensional
PS density, $f_{eq}$, yield the generalized 4-dimensional flux of
particles with momentum
$\vec{p}$:
$ \ \ f_{eq}(p,x)\ \frac{p^\mu}{p^0} . $

Here $G$ is a Gaussian time distribution or emission density for time-like
directed fronts:
$$
   d\tau \ G(\tau - \tau_{FO}) = \frac{d\tau}{\sqrt{2\pi}\ \tau_{coll.}}
   \exp\left[ - \frac{(\tau - \tau_{FO})^2}{2 \tau^2_{coll.}}\right]
$$
and for space-like fronts:
$$
   ds \ G(s - s_{FO}) = \frac{d s}{\sqrt{2\pi}\ \lambda_{m.f.p.}}
   \exp\left[ - \frac{(s - s_{FO})^2}{2 \lambda^2_{m.f.p.}}\right]
$$ .

In more complex models the emission probabilities may take more realistic and
more complicated forms. Furthermore, $f_{e.q.}$ can be space-time dependent,
and can be determined {\it self-consistently} during the detonation,
deflagration or freeze out process \cite{An99a,Ma99a,Ms03a,Mo03a} .

Let us rewrite the escape probability, (\ref{EmiProb}), with the hyper-surface
element,
covering both the timelike and spacelike parts of the freeze-out surface,
\be
  P^*_{esc-PS} =
   \frac{p^{\mu}d^3\sigma_{\mu}}{(p^{\nu}u_{\nu})}\
   \Theta(p^{\mu}d^3\sigma_{\mu}).
    \label{esc1}
\ee
Here $d^3\sigma_{\mu}(x)$ and $u_{\mu}(x)$ are ST dependent and this yields
the secondary $x-$dependence of $P^*_{esc-PS}(p;x)$.
If we take the four velocity equal to $u_{\mu}=(1,0,0,0)$, in the
Rest Frame of the Front (RFF), i.e. where $d^3\sigma_{\mu} = (1,0,0,0)$,  then
$P$ is unity. Otherwise, in the Rest Frame of the Gas (RFG), where $u_{\mu}=(1,0,0,0)$, the escape probability $P$ is
$
    P^*_{esc-PS} =
    {p^{\mu}d^3\sigma_{\mu}} \ \Theta(p^{\mu}d^3\sigma_{\mu})/{p^{0}}.
$

\begin{figure}[!htb]
\centering
\includegraphics[width=8cm, height = 6.5cm]{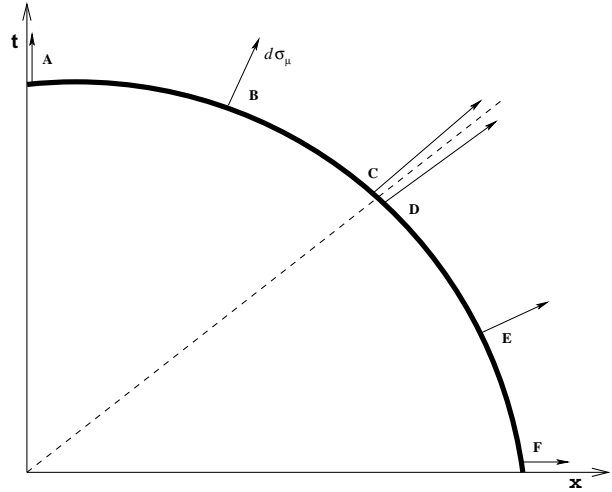}
\caption{
A simple FO-surface in the Rest Frame of the Gas (RFG), where
$u^{\mu} = (1,0,0,0)$, including time-like and space-like parts.
Then the normal vector of the FO front, $d^3\sigma_{\mu}(x)$,
is time-like at the
time-like part and it is changing smoothly into a space-like 4-vector in
the space-like part. On these two parts of the hyper-surface,
in the Local Rest Frame of the Front (RFF), $d^3\sigma_{\mu}(x)$ points into the
direction of the $t'$ ($x'$) axis respectively.
}
\label{updatedFO}
\end{figure}

The nominator depends on $d^3\sigma_{\mu}$. In the following, we will take different typical values for $d^3\sigma_{\mu}$,
for characteristic regions (A, B, C, D, E, F) of the
freeze-out hyper-surface (see Figure \ref{updatedFO}):
\begin{itemize}
\item A
      \ $d^3\sigma_{\mu} = (1,0,0,0)$, leads to
      $p^{\mu}d^3\sigma_{\mu} = p^0 \ge 0 $, as we have seen above
      for FO along the time axis in RFG,
\item B
      \ $d^3\sigma_{\mu} = \gamma (a,b,0,0)$,
\item C
      \ $d^3\sigma_{\mu} = \gamma (1+\epsilon,1-\epsilon,0,0)$,
      just above the light-cone,
\item D
      \ $d^3\sigma_{\mu} = \gamma (1-\epsilon,1+\epsilon,0,0)$,
      just below the light-cone,
\item E
      \ $d^3\sigma_{\mu} = \gamma (c,d,0,0)$,
\item F
      \ $d^3\sigma_{\mu} = (0,1,0,0)$,
      leads to $p^{\mu}d^3\sigma_{\mu} = p^x $,
      for FO along the spatial axis $x$ in RFG,
\end{itemize}
where the $\gamma$ factors serve the unit normalization
of the normal vectors (see Figure \ref{updatedFO}).
The effect of this relativistically invariant FO factor
leads to a smoothly changing behaviour
as the direction of the normal vector changes
in RFG (see Figure \ref{updatedFO}).

\begin{figure}[!htb]
\centering
\includegraphics[width=12cm, height = 9cm]{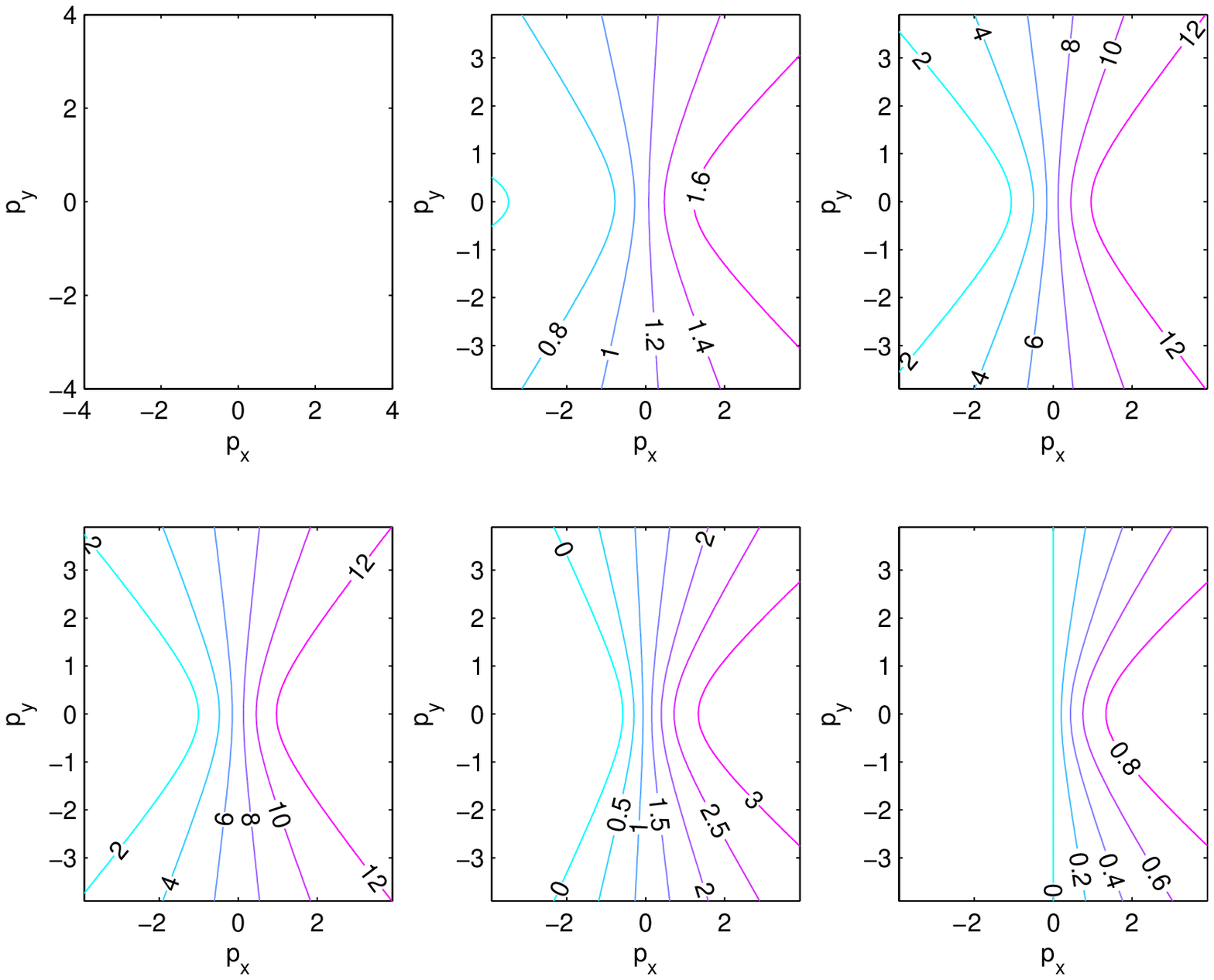}
\caption{
The contour plots of the escape probability factor as function
of the particle momentum ,
$P(\vec{p})$,
at different points of the FO hyper-surface are presented in
six subplots. All plots are in the Rest Frame of the Gas
($RFG$).
For region $A$: $d^3{\sigma}_\mu = (1,0,0,0)$,
and the probability factor is one uniformly,
for $B$:
$d^3{\sigma}_\mu = \gamma_{\sigma} (1,0.5,0,0)$, $C$
(first row) $D$, $E$:
$d^3{\sigma}_\mu = \gamma_{\sigma} (1,2,0,0)$
$F$: $d^3{\sigma}_\mu = (0,1,0,0)$ (second row).
The momenta are in units of $[mc]$.}
\label{P_B_eset}
\end{figure}

To calculate the parameters of the normal vector $d^3\sigma_{\mu}$
for different cases listed above, we simply make use
of the Lorentz transformation. The normal vector
of the time-like part of the freeze-out hypersurface may be
defined as the local $t'$-axis, while the normal vector for
the space-like part may be defined as the local $x'$-axis.
As the $d^3{\sigma}_\mu$ normal vector is normalized to unity
its components may be interpreted in terms of
$\gamma_{\sigma}$ and $v_{\sigma}$, as
$d^3\sigma_{\mu}=\gamma_{\sigma} (1,v_{\sigma},0,0)$,
where
$\gamma_{\sigma} = \frac{1}{\sqrt{1 - v_{\sigma}^2}}$
for time-like normals and
$\gamma_{\sigma} = \frac{1}{\sqrt{v_{\sigma}^2-1}}$
for space-like normals.
This will lead to,

\begin{itemize}
\item A \ $d^3\sigma_{\mu} = (1,0,0,0)$, leads to $P = 1$,
\item B \ $d^3\sigma_{\mu} = \gamma_{\sigma} (1,v_{\sigma},0,0)$, leads
          to $P= \frac{\gamma_{\sigma} ( p^0 + v_{\sigma} p^x)}{p^0}$,
      and $\gamma_\sigma = \frac{1}{\sqrt{1-v^2_{\sigma}}}$,
\item C \ $d^3\sigma_{\mu} = \gamma_{\sigma} (1+\epsilon,1-\epsilon,0,0)$,
          leads to $P = \frac{4 \gamma_{\sigma}^2(p^0 + p^x) +
      \gamma_{\sigma}(p^0 + p^x)}{4 \gamma_{\sigma} p^0}  $,
\item D \ $d^3\sigma_{\mu} = \gamma_{\sigma} (1-\epsilon,1+\epsilon,0,0)$,\\
          leads to $P = \frac{4 \gamma_{\sigma}^2(p^0 + p^x) -
      \gamma_{\sigma}(p^0 - p^x)}{4 \gamma_{\sigma} p^0}
      \times  \Theta(p^{\mu}d^3\sigma_{\mu})$,
\item E \ $d^3\sigma_{\mu} = \gamma_{\sigma} (1,v_{\sigma},0,0)$,\\
          leads to $P= \frac{\gamma_{\sigma} ( p^0 + v_{\sigma} p^x)}{p^0}
      \times  \Theta(p^{\mu}d^3\sigma_{\mu})$,
      and $\gamma_\sigma = \frac{1}{\sqrt{v^2_{\sigma}-1}}$,
\item F \ $d^3\sigma_{\mu} = (0,1,0,0)$, leads to $P = \frac{p^x}{p^0}
      \times  \Theta(p^{\mu}d^3\sigma_{\mu})$,
          (see Figure \ref{updatedFO}).
\end{itemize}

The resulting Phase Space escape probabilities are shown in Figure
\ref{P_B_eset} for the six cases described above.

In refs. \cite{An99a,Ma99a} the post FO distribution was evaluated for
space-like gradual FO in a kinetic model. Initially we had an equilibrated,
interacting, PS distribution, $f_{int}(p,x)$, and an escape probability,
similar to eq. (\ref{esc1}), but simplified so that it depended on the
angle of the two vectors only. After some small fraction of particles
were frozen out as the FO process progressed in the front, the interacting
component were re-equilibrated, with smaller particle number, smaller
energy and momentum, to account for the quantities carried away by the
frozen out particles. This was then repeated many times in small steps
along the FO front and the frozen out particles were accumulated in the
post freeze out PS distribution, $f_{free}$. The resulting distribution
was highly anisotropic and obviously non-equilibrated. The details of the
post FO distribution depend on the details of the escape probability, and
on the level of re-equilibration of the remaining, interacting component.

\begin{figure}[!htb]
\centering
\includegraphics[width=8cm, height = 12cm, angle=-90 ]{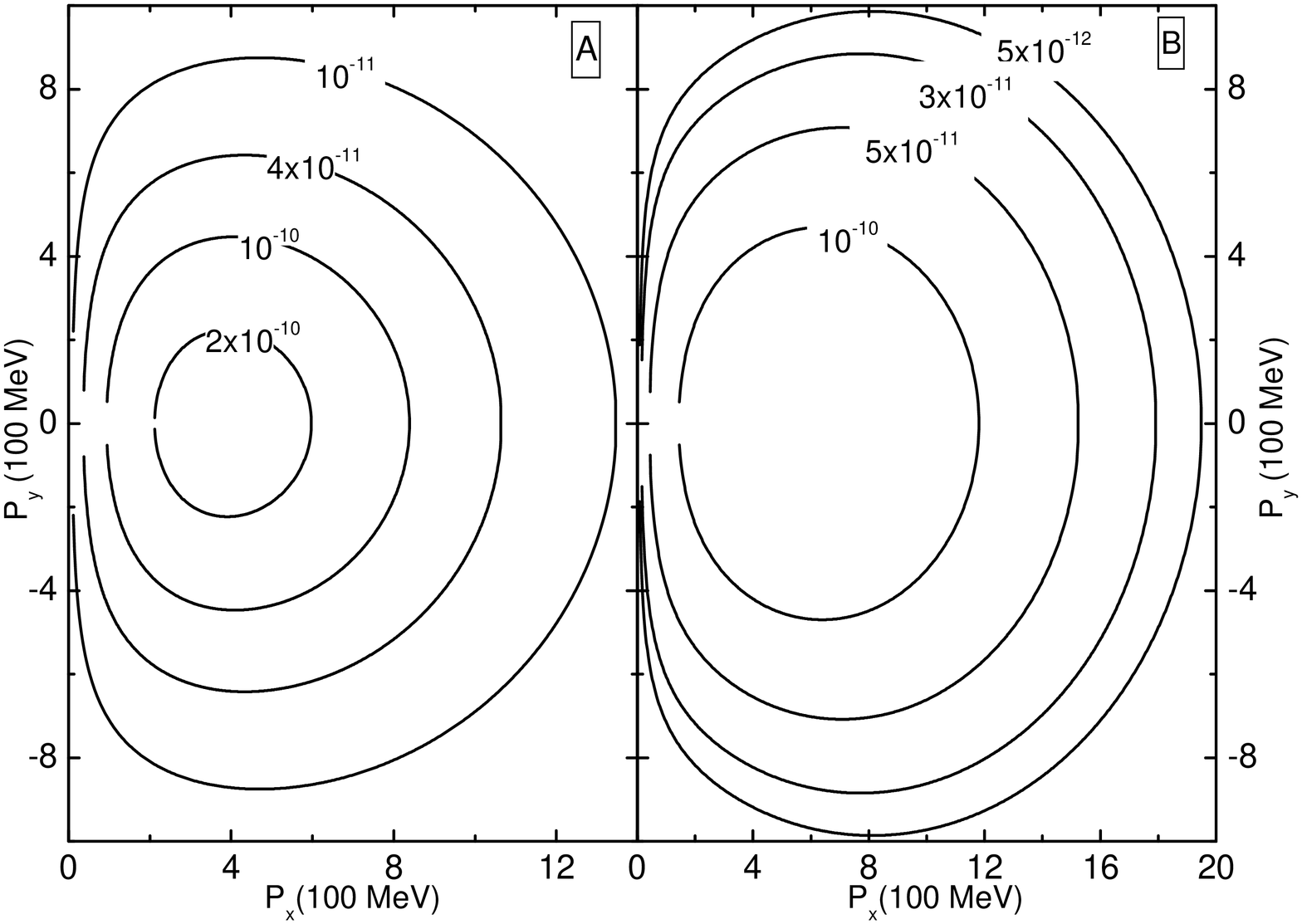}
\caption{
A simple FO-surface in the Rest Frame of the Gas (RFG) where
$d^3\sigma_{\mu} = (1,0,0,0)$,
including timelike and spacelike parts.
Then $d^3\sigma_{\mu}$ is $t' = d^3\sigma_{\mu}$
for the timelike part, changing smoothly into
$x' = d^3\sigma_{\mu}$
for spacelike normal vectors.}
\label{Karolis3-f2}
\end{figure}

Bugaev assumed earlier \cite{Bu96}, that the post FO distribution is a
(sharply) "Cut-Juttner" distribution, but the above mentioned model shows
that this can only be obtained if re-equilibration is not taking place.
The kinetic model provided an asymmetric but smooth PS distribution
\cite{An99a}, while the escape probability (\ref{esc1}) yields a somewhat
different, but also smooth PS distribution. These can be well approximated
by the "Cancelling Juttner" distribution \cite{Ta03},
Figure \ref{Karolis3-f2}.

Up to now we analysed the FO probability in the momentum space only.
At this point we should also mention, that if we discuss gradual FO across a
finite layer, the distance from a {\it hypothetical} or {\it theoretical}
FO surface must also be part of the FO probability, because the
probability of a particle has no more collision on its way out depends
on the local mean free path (m.f.p.) and on the distance from this
surface. The m.f.p. is usually momentum dependent and this yields the
secondary $p-$dependence of $P_{esc-ST}(x;p)$.
Of course the position of such a {\it hypothetical} FO surface
has to be determined self consistently or within a realistic kinetic
model.  However, if we assume simultaneous FO and hadronization
\cite{CC94,CM95} from
supercooled plasma, there are no realistic and reliable dynamical models
up to know, which we could use, so, a now the first self consistent approximate
method is our only choice.

\subsection{Emission on an idealized hyper-surface}

If the m.f.p. or the coll. time are negligibly small compared to the
ST dimensions of the system we can idealize the source distribution further
as the source is shrinking to a hyper-surface (HS) in the ST, i.e.:
$$
   \frac{d \tau}{\sqrt{2\pi}\ \tau_{coll.}}
   \exp\left[ - \frac{(\tau - \tau_{FO})^2}{2 \tau^2_{coll.}}\right]
   \longrightarrow  \delta (\tau-\tau_{FO})\ d\tau\ {\rm : time-like\ fronts,}
$$
$$
   \frac{d s}{\sqrt{2\pi}\ \lambda_{m.f.p.}}
   \exp\left[ - \frac{(s - s_{FO})^2}{2 \lambda^2_{m.f.p.}}\right]
   \longrightarrow  \delta (s - s_{FO})\ ds\  {\rm : space-like\ fronts,}
$$
and the integration in the normal direction to the front can be
executed. Then
\be
   N_{new} = \int \frac{d^3p}{p^0} \ \int_\Sigma \ (p^\mu d^3\sigma_\mu)\
   \Theta(p^\mu\ d^3\sigma_\mu)\   f_{FO}(p,x)\   C_1 \ .
   \label{eq21}
\ee
Here $_\Sigma$ denotes integration over a 3-dimensional HS.
If all particles described by $f_{e.q.}$ are emitted during the
process, $C_1 = 1$. If we do not perform the momentum
integral we recover the modified "Cooper-Frye" freeze out
formula\cite{CF74,Bu96,ACG99}
\be
   p^0 \frac{dN}{d^3p}   =    \int_\Sigma  p^\mu d^3\sigma_\mu \
   \Theta(p^\mu\ d^3\sigma_\mu)\  f_{FO}(p,x)
   \label{CFformula}
\ee
Notice that we have changed the notation of the PS distribution
from  $f_{eq}$ to $f_{FO}$ denoting the post FO distribution
integrated across the front.

Eq. (\ref{CFformula}) simplifies the process to a great extent,
especially if we consider that $f_{eq}$ changes during the detonation
deflagration or freeze out process as we go across the front
self-consistently. In a dynamical process across the front,
even if $f_{eq}$ is always an
equilibrated distribution, its parameters change in a wide range
\cite{An99a,Ma99a}.
In the FO process the post FO distribution, after the
process is completed is very far from any statistically equilibrated
distribution. This should be taken into account if we apply the assumption of
an idealized FO surface.
The determination of the FO surface is an involved task, both physically and
technically. In complex modeling tasks we perform the self-consistent
determination of the surface in simplified models. In these models we can
pinpoint the FO conditions and evaluate the post FO distributions.

Then based on this knowledge we determine the FO surface in the
large scale, 3+1 dimensional system by analysing the complete ST history
of the fluid dynamical stage calculated. This demanding task was completed
recently by {\it Bernd R. Schlei} (of Los Alamos) \cite{SchleiIP},
and this enables us to visualize
the ST evolution of the surface where the emitted particles originate
from.

\section{Two particle correlations}

In case of two particle correlations we have to return to
eqs. (\ref{NewH}) or (\ref{eq11}), because the emission point in ST
is crucial. Although, eqs. (\ref{eq21}) or (\ref{CFformula}) also
contain a ST dependence, this is reflecting the properties of the
idealized hypersurface, where the process is taking place. This
approximation is justified only if
\begin{itemize}
\item The ST domain of the emission is a negligibly narrow layer
      and we are sure that the internal processes within the front
      are not relevant.
\item If we use a realistic estimate for the post FO distribution
in these equations.
\end{itemize}

The PS density of created (or frozen out) particles, $N_{cr}$, is given
in terms of a "Source function", $S(x,p)$, as \cite{Sch92}
\footnote{Note that in some cases the source function is defined differently
\cite{Chapm}, so that the $p^0$ factor is not included in the denominator:
$ p^0 \ \frac{d N_{cr}}{d^3p} = \int S'(x,p) d^4x\ $.}
\be
\frac{d N_{cr}}{d^3p} = \int S(x,p)\ d^4x\ ,
\ee

which can be compared to eqs. (\ref{NewH},\ref{esc1}), so that
\be
p^0\  \frac{d N_{cr}}{d^3p} =
\int d^4x\ f_{eq}(p,x)\ \ P_{esc-PH}(p;x)\ \ P_{esc-ST}(x;p) \ \ {\rm or}
\ee
\be
 \frac{d N_{cr}}{d^3p} =
\int d^4x\
\underbrace{f_{eq}(p,x)\ \ \overbrace{P_{esc-PH}(p;x)/p^0}^{P^*_{esc-PH}(p;x)}
\ \ P_{esc-ST}(x;p)}_{S(p;x)} \ .
\ee

Thus, assuming emission or freeze-out from an interacting gas the source
function is
\be
S(x,p) = f_{eq}(p,x)\ P^*_{esc-PH}(p;x)\ P_{esc-ST}(x;p)\ ,
\ee
here we assumed that the emission or escape probability is factored to
a PS and a ST dependent part.

Two particle correlations provide information about the location of FO points
in the ST, providing the possibility of deeper insight into the
reaction mechanism.  This obviously means that there is an essential difference
between reaction models assuming (i) FO in an extended ST layer or (ii)
FO accross an idealized ST hypersurface.  The latter assumption is
obviously not realistic.

Nevertheless, the idealized hypersurface assumption may not be so bad as one
would think. Even in this case parts of this surface can represent early times
while others late emissions. The spatial locations of different parts
of the surface can be far from each other. If the ST dimensions of the
idealized FO hypersurface are much larger than the thickness
of the real FO layer, the hypersurface assumption may be satisfactory.
In any case, two particle correlations are much more sensitive to this
simplifying assumption than single particle data.

Note also that the thickness of the FO layer is dependent of the
hadronic species measured.  Thus, the hypersurface idealization
is applicable for high multiplicity hadrons, protons and pions, while
particles with low multiplicity, low cross-sections or low creation rates
are not described well with this approximation.

\section{Simultaneous Hadronization and Freeze-Out}

If one assumes
hadronization in thermal and fluid dynamical equilibrium via homogeneous
nucleation \cite{CK92} or similar processes, this leads to a
lengthy, gradual hadronization and freeze-out. Such a long-lived particle
source should be detected by pion interferometry, but
recent data from RHIC ruled out this scenario.
Therefore we do not assume a lengthy thermalization and
chemical equilibration process at freeze-out either.
As we described in ref. \cite{KCM03}, we assume simultaneous FO and
hadronization at the end of the heavy ion reaction.

We assume that pre-hadrons or quasi-hadrons are formed in the
cooling and expanding plasma before we reach the critical temperature
and density, and especially when the plasma supercooles. Supercooling is
possible even if the phase transition is just a smooth, {\it but sharp}
cross-over, which is always the case in small, finite systems. Then
hadrons are formed out of thermal equilibrium, which are frozen out
(i.e. never collide) after their formation.  No detailed hadronization
models exist for this kind of mechanism. The closest are coalescence or
recombination models, which usually describe the hadron abundances well.

As discussed in \cite{KCM03} and \cite{Cs02}, the similarity of the
results arising from statistical (thermal) and coalescence models,
is due to the fact that the Canonical Ensemble (CE) is generated if the
average energy of the system is the same in the elements of the ensemble
and the states included are generated with equal a priory probability.
For most hadrons these conditions are satisfied, so the two models yield
similar results and therefore we use the statistical model to estimate
local hadron abundances at the end of the reaction \cite{KCM03}.

The differences arise for hadrons where the assumption of equal a priory
formation probability does not hold. An example is the $\Lambda(1520)$
resonance, an excited state where the radial wave function includes both
s- and p- waves, and so the formation cross-section is much less than for
other hadrons. Therefore the abundance of these hadrons is much smaller, and
this is well described by coalescence type models, while statistical
models fail to reproduce the smaller abundance (as the cross-section does not
appear in these models).

For our purposes, however, the statistical model is adequate, because
it includes all necessary statistical weights, (so it works well for
the high multiplicity hadrons,) while low multiplicity resonances are
not reproduced by fluid dynamical models anyway.

\section{Summary}

Recent experimental data indicate that detailed fluid dynamical data
are becoming available at RHIC energies. This will provide us with the
possibility to test the QGP Equation of State in heavy ion reactions.
This will require further experimental efforts that provide us with
the complete spectrum of collective flow data. To draw the right conclusion
will also require a detailed and realistic theoretical reaction model
which can simultaneously describe single particle and two-particle
observables.

%

\bibliographystyle{amsunsrt}

\vfill
\end{document}